\documentclass[10pt]{article}
\usepackage[margin=1in]{geometry}                
\usepackage{graphicx}
\usepackage{amssymb}
\usepackage{epstopdf}
\usepackage{amsmath}
\usepackage{hyperref}
\usepackage{mathrsfs}
\usepackage{varioref}
\usepackage{textcomp}
\usepackage{enumerate}
\usepackage{paralist}
\usepackage{comment}
\usepackage{titlesec}
\usepackage{cancel}

\hypersetup{
    bookmarks=false,         
    pdfstartview={FitH},    
    colorlinks=true,       
    linkcolor=black,          
    citecolor=black,        
    filecolor=blue,      
    urlcolor=blue           
}

\DeclareFontFamily{OT1}{pzc}{}
\DeclareFontShape{OT1}{pzc}{m}{it}%
             {<-> s * [0.900] pzcmi7t}{}
\DeclareMathAlphabet{\mathscr}{OT1}{pzc}%
                                 {m}{it}

\titleclass{\subsubsubsection}{straight}[\subsection]

\newcounter{subsubsubsection}[subsubsection]
\renewcommand\thesubsubsubsection{\thesubsubsection.\arabic{subsubsubsection}}

\titleformat{\subsubsubsection}
  {\normalfont\normalsize\bfseries}{\thesubsubsubsection}{1em}{}
\titlespacing*{\subsubsubsection}
{0pt}{3.25ex plus 1ex minus .2ex}{1.5ex plus .2ex}

\makeatletter
\renewcommand\paragraph{\@startsection{paragraph}{5}{\z@}%
  {3.25ex \@plus1ex \@minus.2ex}%
  {-1em}%
  {\normalfont\normalsize\bfseries}}
\renewcommand\subparagraph{\@startsection{subparagraph}{6}{\parindent}%
  {3.25ex \@plus1ex \@minus .2ex}%
  {-1em}%
  {\normalfont\normalsize\bfseries}}
\def\toclevel@subsubsubsection{4}
\def\toclevel@paragraph{5}
\def\toclevel@paragraph{6}
\def\l@subsubsubsection{\@dottedtocline{4}{7em}{4em}}
\def\l@paragraph{\@dottedtocline{5}{10em}{5em}}
\def\l@subparagraph{\@dottedtocline{6}{14em}{6em}}
\makeatother

\labelformat{section}{Section #1} 
\labelformat{subsection}{Section #1} 
\labelformat{subsubsection}{Section #1}
\labelformat{subsubsubsection}{Section #1}
\labelformat{equation}{Eq.~(#1)} 
\labelformat{figure}{Fig.~#1} 
\labelformat{subfigure}{Fig.~\thefigure#1} 
\labelformat{table}{Tab.~#1} 
\newcommand{\be}{\begin{equation}}
\newcommand{\ee}{\end{equation}}
\newcommand{\bea}{\begin{eqnarray}}
\newcommand{\eea}{\end{eqnarray}}
\newcommand{\nn}{\nonumber}

\newcommand{\sqg}{\sqrt{-g}}
\newcommand{\sqh}{\sqrt{|h|}}

\newcommand{\mLs}{\mathcal{L_{\textrm{sur}}}}

\newcommand{\df}{\delta}

\newcommand{\intdx}{\int_\mathcal{\partial V} d^3x}

\newcommand{\ST}{\df \mathcal{A}_{_{\partial  \mathcal \tiny V}}}

\newcommand{\us}{u_{(s)}}

\newcommand{\CO}{\mathcal{C}_{0}}
\newcommand{\CG}{\mathcal{C}_{\textrm{GHY}}}
\newcommand{\CE}{\mathcal{C}_{E}}

\def\m#1{\mathcal{#1}}

\def\P[#1]#2{\Pi^{#1}_{\phantom{a}#2}}

\includecomment{comment}

\begin{document}
	\title{Variational Principle for Gravity with Null and Non-null boundaries: A Unified Boundary Counter-term}
	\author{ 
		{\bf {\normalsize Krishnamohan Parattu$^a$,}$
			$\thanks{E-mail: krishna@iucaa.ernet.in}} \,
		{\bf {\normalsize Sumanta Chakraborty$^a$,}$
			$\thanks{E-mail: sumanta@iucaa.in}} \, \\\\ and
		{\bf {\normalsize T. Padmanabhan$^a$}$
			$\thanks{E-mail: paddy@iucaa.in}}\\\\
		{\normalsize $^a$IUCAA, Post Bag 4, Ganeshkhind,}
		\\{\normalsize Pune University Campus, Pune 411 007, India}
		\\[0.3cm]
	}\date{\today}
	\maketitle
	
	\begin{abstract}
		It is common knowledge that the Einstein-Hilbert action does not furnish a well-posed variational principle. The usual solution to this problem is to add an extra boundary term to the action, called a counter-term, so that the variational principle becomes well-posed. When the boundary is spacelike or timelike, the Gibbons-Hawking-York counter-term is the most widely used. For null boundaries, we had proposed a counter-term in a previous paper. In this paper, we extend the previous analysis and propose a counter-term that can be used to eliminate variations of the ``off-the-surface" derivatives of the metric on any boundary, regardless of its spacelike, timelike or null nature.
	\end{abstract}
	
	

	\section{Introduction}
	An action for general relativity was introduced very early in the history of general relativity, back in 1915, by David Hilbert, and is now known as the Einstein-Hilbert action. The dynamical variable involved is the metric. Einstein-Hilbert action is unlike the usual actions found in field theory since it contains the second derivatives of the metric. But the Einstein equations obtained from the action are only second order in the derivatives of the metric. Hence, the theory does not have the usual problems associated with higher order field equations. The place where the second derivatives in the action make their presence felt is in the boundary term in the variation of the action, which requires fixing both the field variable (the metric) and its normal derivatives at the boundary. Requiring such boundary conditions, in conjunction with the second order equations of motion, will lead to a variational principle that is not well-posed \cite{Dyer:2008hb}.
	
	The standard solution is to add a counter-term to the action to produce a combined action which would require only the metric to be fixed at the boundary. The most popular counter-term used is the Gibbons-Hawking-York (GHY) counter-term \cite{York:1972sj, Gibbons:1976ue}, but it is by no means unique \cite{Charap:1982kn}. In fact, Einstein himself is perhaps the first person to use this technique \cite{Einstein:1916cd}, although he would have probably phrased it as throwing away a total divergence term from the action (rather than adding a counter-term).
	
	The GHY term is constructed using the unit normal to the boundary surface and hence is not directly applicable to a null surface (unless one treats the null surface as the limit of non-null surfaces). We addressed this issue in \cite{Parattu:2015gga}, where we proposed a covariant counter-term that can be used for a null surface. The natural question that arises is whether we can find a common prescription that can be used without specifying whether the surface is null or non-null. We answer this question in this work and provide such a prescription. We shall also show how our prescription relates to the other prescriptions mentioned above.
	
	The conventions used in this paper are as follows: We use the metric signature $(-,+,+,+)$. The fundamental constants $G$, $\hbar$ and $c$ have been set to unity. The Latin indices, $a,b,\ldots$, run over all space-time indices, and are hence summed over four values. Greek indices, $\alpha ,\beta ,\ldots$, are used when we specialize to indices corresponding to a codimension-1 surface, i.e a $3$-surface, and are summed over three values. Upper case Latin symbols, $A,B,\ldots$, are used for indices corresponding to two-dimensional hypersurfaces, leading to sums going over two values. For a $\phi=\textrm{constant}$ surface, $s_c$ is used to refer to the surface gradient, $\nabla_c \phi$, and $v_c$ is used to refer to a normal vector with an arbitrary normalization factor (i.e. $v_c=A\nabla_c \phi$ for arbitrary $A(x^a)$). For non-null surfaces, $n_c$ is the normalized normal. For null surfaces, we shall use $\ell_c$ as the normal, which we shall take to be equal to the surface gradient $s_c$ since we do not know of any other natural normalization. The auxiliary vector is in general represented by $t^c$, while $k^c$ will be used when we specialize to an auxiliary null vector for a null boundary. 
	
	\section{Boundary Terms in General Relativity: An Overview} \label{sec:Action_Overview}
	Consider a four-dimensional spacetime volume $\m{V}$ with a three-dimensional boundary $\partial\m{V}$. Introduce a scalar function $\phi(x^a)$ such that it is constant on the boundary and define the ``surface gradient" $s_c\equiv\nabla_c \phi$. Variation of the Einstein-Hilbert action in $\m{V}$ will then generate the following contribution on the boundary $\partial \m{V}$ \cite{gravitation}:
	\begin{equation}
	16\pi \left[\ST \right]_{\rm EH}= \int_\mathcal{V}  d^4x  \,\, \partial_c\left[\sqg  \left(g^{ab} \delta \Gamma^{c}_{ab}-g^{ck} \delta \Gamma^{a}_{ak}\right)\right]= \int_\mathcal{\partial V}  d^3x \sqg\,\, \epsilon s_c  \left(g^{ab} \delta \Gamma^{c}_{ab}-g^{ck} \delta \Gamma^{a}_{ak}\right)~.
	\end{equation}
	Here, $\epsilon$ is $+1$ or $-1$ as per the conventions of Gauss' theorem. Since this is an overall constant factor, we shall not mention this factor in our future manipulations. But due care must be taken to include it in any specific applications.
	Defining $v_c\equiv A \partial_c \phi$, with an arbitrary normalization factor $A(x^a)$, and $Q[X_c]$ for any one-form $X_c$ as 
	\begin{equation}\label{Q_def}
	Q[X_{c}] \equiv X_{c} (g^{ab} \delta \Gamma^{c}_{ab}-g^{ck} \delta \Gamma^{a}_{ak})~,
	\end{equation}
	the boundary term becomes
	\begin{equation} \label{ST_in_Q}
	16\pi \left[\ST \right]_{\rm EH} = \int_\mathcal{\partial V}  d^3x \frac{\sqg}{A}\, Q[v_c]~. 
	\end{equation}
	For example, we can take $\phi$ to be $t$ and $v_a$ to be the normalized normal $n_a=N\partial_a t$ for a spacelike boundary in ADM formulation \cite{Arnowitt:1962hi}, where $N$ is the lapse function. Then, the integration measure in the above equation becomes the familiar $d^3x \sqh$, with $h$ the determinant of the $3$-metric on the surface.
	   
	   To set \ref{ST_in_Q} to zero, it is not enough to fix the metric at the boundary; we need to fix the derivatives of the metric as well. Such a structure arises because of the presence of second derivatives in the Einstein-Hilbert action. The \textit{tangential} derivatives get fixed automatically when we fix the metric at the boundary. But the variations of the \textit{normal} derivatives of the metric have to be set to zero separately. The difficulty is that fixing both the metric and its normal derivatives at the boundary may not provide a consistent solution to the variational problem \cite{Dyer:2008hb,Parattu:2013gwb}.
	The commonly accepted solution is to modify the action by adding an extra boundary term, called the counter-term, to the action to remove the variation of the normal derivatives. The counter-term that can be added is not unique \cite{Charap:1982kn}. We shall next discuss some counter-terms that are available in the literature.
	\subsection{Counter-terms to the Einstein-Hilbert Action}
	\subsubsection{The Einstein Counter-term}
	Einstein separated out a total divergence term, $\mLs$, from the Einstein-Hilbert Lagrangian density, $\mathcal{L}_{\rm EH}$, such that it contains all the second derivatives of the metric \cite{Einstein:1916cd}, and worked with the remaining quadratic (also known as bulk or $\Gamma^2$) Lagrangian density $\mathcal{L}_{\rm quad}$. The total divergence term is
	\begin{align}
		\mLs &=  \partial_c \left(\sqg V^c \right); \quad V^c= -\frac{1}{g} \partial_b \left(g g^{bc} \right)~.
	\end{align}
In our terminology, Einstein added $-\partial_c \left(\sqg V^c \right)$ as a counter-term to the Lagrangian. Using Gauss's theorem, we will define the Einstein counter-term $\mathcal{C}_{E}$ as the integral over the boundary surface of the quantity
	\begin{equation}\label{CTE1}
		\mathcal{C}_{E}\equiv - \sqg s_c V^c = -\sqh n_c V^c ~,  
	\end{equation}
	where the second equality is valid for a non-null surface, with $n_c$ the unit normal and $h$ the determinant of the surface metric.
\subsubsection{The Gibbons-Hawking-York (GHY) Counter-term}
	The expression for the Einstein counter-term is not covariant. A covariant counter-term, that is most popular at present and is part of the newer textbooks \cite{gravitation,Wald,Poisson}, is the Gibbons-Hawking-York counter-term \cite{York:1972sj, Gibbons:1976ue}. It is the integral over the boundary surface of
	\begin{equation}\label{CTGHY1}
	\mathcal{C}_{\textrm{GHY}}\equiv -2 \sqh K=2 \sqh \nabla_a n^a~.
	\end{equation}
	But note that this covariance is achieved by the introduction of the normal one-form to the boundary surface, $n_{a}$, as a variable, in addition to the metric. Since the action is well-defined only when the Lagrangian as well as the region $\mathcal{V}$ are specified, one might think that the normal $n_i$ to the boundary $\partial\mathcal{V}$ is also available to us. But the action in \textit{any} theory requires specification of $\mathcal{V}$ for its definition; but we never need $n_i$ to be introduced into the action in any other theory; in this sense, gravity \textit{is} somewhat special. This uniqueness originates from the fact that we need to include second derivatives of the metric in the action for invariance under coordinate transformations, unlike in other field theories. It should be stressed that $\CE$ and $\CG$ are not equal except when the shift functions $N^{\alpha}$ are constants on the boundary (see Chapter 6 in \cite{gravitation}), but their variations are equal when the metric components, in particular $N^{\alpha}$, are fixed on the boundary.  We shall explicitly show the difference between the various counter-terms under consideration in \ref{CT_compare}. 
	
	\subsubsection{A Counter-term for Null Surfaces}
	The GHY prescription has the disadvantage that it applies directly only to a non-null surface. This is because $K$ diverges, as the normalization of the unit normal diverges on a null surface, while $\sqrt{h}$ goes to zero. By a proper limiting procedure, it can be proved that $\sqrt{h}K$ has a finite limit. Still, it is cumbersome to use this expression on a null surface. In \cite{Parattu:2015gga}, we proposed a counter-term for a null boundary that is constructed from quantities well-defined on a null surface. It is the integral over the boundary of 
	\begin{equation}\label{CT01}
		\mathcal{C}_{n}\equiv 2 \sqrt{q} \left(\Theta + \kappa\right),
	\end{equation}
	where $q$ is the determinant of the $2$-metric on the null surface, $\Theta$ is the expansion of the null geodesics on the null surface, and $\kappa$ is the surface gravity.
\section{Boundary Term for a General Surface}
A natural question to ask is if we can get rid of the division into null and non-null and propose a covariant counter-term that can be used on any type of boundary. For a non-null surface, we require a normalized normal, of which there seems to be no well-defined notion on a null surface. Thus, we shall attempt to extend the considerations that led us to a counter-term for a null surface to the case of a non-null surface.

In a $4$-dimensional spacetime manifold $M$, consider a $4$-dimensional region $\mathcal{V}$ with a $3$-dimensional boundary surface denoted by $\partial \mathcal{V}$. Introduce a scalar $\phi$ such that $\partial \mathcal{V}$ is represented by the equation $\phi=\phi_0$ for some constant $\phi_0$. We shall often find it useful to evaluate expressions in a ``3+1" coordinate system with $\phi$ as a coordinate, i.e
\begin{equation} \label{spec_coord}
x^{a}=(\phi,y^1,y^2,y^3),
\end{equation} 
where $(y^1,y^2,y^3)$ are arbitrary. Then, Greek indices will be used to run over $(y^1,y^2,y^3)$. In particular, this allows the separation of normal derivatives ($\partial_\phi$) and surface derivatives ($\partial_\alpha$). The normal one-form $v_a$ to the surface $\partial \mathcal{V}$ is defined by $v_a= A \nabla_a \phi$, where $A(x^i)$ is some scalar function which is finite and non-zero on $\partial \mathcal{V}$. Choose an auxiliary vector $t^a$ such that $t^a v_a=-1$. This condition ensures that the vector $t^a$ does not lie on the surface. Hence, three basis vectors on the surface and $t^a$ form a basis for the four-dimensional spacetime near the boundary. In this basis, we can project any vector to the surface by removing its component along $t^a$. For a non-null surface, $v^a$ is not on the surface and hence can be normalized to form $t^a$. For a null surface, $v^a$ is on the surface and hence $t^a$ has to be chosen to be linearly independent from $v^a$. We can now form the mixed tensor 
\begin{equation}\label{P}
\Pi^a_{\phantom{b}b}=\df^a_b+t^a v_b~.
\end{equation}
This is a projector of vectors onto the tangent space of the $\phi$-constant surfaces ($B_{\perp}^a=\P[a]{b}B^{b}\Rightarrow B_{\perp}^a v_a=0$ and $\Pi^{a}_{\phantom{b}b}\Pi^{b}_{\phantom{b}c}=\Pi^{a}_{\phantom{b}c}$). In terms of the basis mentioned above, it removes the component along $t^b$, as $\Pi^a_{\phantom{b}b}t^b=0$, while leaving the components along the basis vectors on the surface intact.

Introduce a metric $g_{ab}$ on $\m{V}$. For the purpose of this paper, we shall assume that the metric is non-degenerate, i.e its determinant is never zero, so that the inverse of the metric exists everywhere. Then, on a non-null surface, we can choose $t^a= -v^a /v^2$ and reduce $\Pi^a_{\phantom{b}b}$ to $h^a_b$ (see also  Carter in \cite{Hawking:2010mca}). But we shall keep $t^a$ arbitrary as far as possible. 

Let us consider the possible choices of scalars $\phi$ and $A$. We shall take $\phi$ to be independent of the metric, so that it need not be varied when the metric is varied. Taking $\phi$ as one of the coordinates, say the time $t$ in a 3+1 split with a spacelike boundary, is an example. For $A$, a natural choice on a non-null surface is the normalization factor. Since we include null surfaces also in our ambit, we cannot make this choice. As there appears to be no other natural choice, we shall put $A=1$ purely for the ease of manipulation. Once we make this choice, $v_a=s_a$, the surface gradient, and $\df v_a=0$. Also, we can no longer identify $v_a$ with a normalized normal except in special cases. But connection to the usual formalism on non-null surface can be made by the choice of $t^a$ discussed above that reduces $\Pi^a_{\phantom{b}b}$ to $h^a_b$.

Varying the Einstein-Hilbert action in the spacetime region $\m{V}$, we obtain \ref{ST_in_Q} as the contribution on the boundary $\partial \m{V}$. 
Our aim is to separate out all the variations of the normal derivatives into a term that can be cancelled by a counter-term, following a procedure that was used to obtain the GHY counter-term in \cite{Padmanabhan:2014BT}. We shall start from the following expression for $\sqg Q[X_c]$:
\begin{equation}
	\sqrt{-g} Q [X_c] = \sqrt{-g} \nabla_c [\delta u_{(X)}^{c}] - 2 \delta (\sqrt{-g} \nabla_a X^{a}) + \sqrt{-g} (\nabla_a X_b -g_{ab} \nabla_c X^{c}) \delta g^{ab}~, \label{qgen} 
\end{equation}
where $\df u_{(X)}^a= \df X^a +  g^{ab} \df X_b $. This expression can be checked by a straightforward evaluation of the RHS. Thus, for the case $v_c=s_c$, the boundary term is the integral on the boundary of
\begin{equation}
\sqrt{-g} Q [s_c] = \sqrt{-g} \nabla_c [\delta u_{(s)}^{c}] - 2 \delta (\sqrt{-g} \nabla_a s^{a}) + \sqrt{-g} (\nabla_a s_b -g_{ab} \nabla_c s^{c}) \delta g^{ab}~.
\end{equation}
Now, $\delta u_{(s)}^a= \df s^a=s_b\df g^{ab}=\df g^{a\phi}$ in our 3+1 coordinates (\ref{spec_coord}).  The second term in \ref{qgen} can be eliminated by adding the integral of $2\sqrt{-g} \nabla_a s^{a}= 2\partial_a \left(\sqg s^a \right)$ over the boundary as a counter-term. But this alone will not be sufficient as the first term does have the normal derivatives of the metric (of form $\sqg \partial_\phi \left(\df g^{\phi\phi}\right)$). This term may be decomposed as follows:
\begin{equation} \label{ST}
\sqrt{-g} \nabla_c [\delta \us^{c}] =\partial_a \left(\sqg \df s^a\right)=\partial_a \left(\sqg \P[a]{b}\df s^b \right)-\partial_a \left(\sqg t^a s_{b}\df s^b \right).
\end{equation}
The first term has only surface derivatives as $\P[\phi]{b}=0$. For the second term, we have
\begin{align} \label{ST2manip}
-\partial_a \left(\sqg t^a s_{b}\df s^b \right)= -\sqg (\nabla_a t^a) s_b \df s^b-2\df\left[\sqg t^a  s_b \nabla_a s^b\right]+2\left[\df\left(\sqg\right) t^a+\sqg \df t^a\right] s_b \nabla_as^b,
\end{align}
where the property $\df s_a=0$ was used.
Using \ref{ST} and \ref{ST2manip} in \ref{qgen}, we obtain
\begin{align} \label{result1}
\sqrt{-g} Q [s_c] &= \partial_a \left(\sqg \P[a]{b}\df s^b \right) - 2 \delta (\sqrt{-g}\P[a]{b} \nabla_a s^{b}) +  \sqg \df t^a \partial_a \left(s^2\right) \nn\\
&+ \sqrt{-g} \left[\nabla_a s_b -g_{ab} \left(\P[c]{d} \nabla_c s^{d}\right)- (\nabla_c t^c)s_a s_b \right] \delta g^{ab}~.
\end{align}
The meaning of the $\df t^a$ term depends on our choice of $t^a$. If $t^a$ is some fixed vector independent of the metric, $\df t^a=0$. If we choose it to be metric-dependent, like the choice $t^a=-s^a/s^2$ we can make for the non-null surface, $\df t^a$ will contain variations of the metric. As long as we do not define $t^a$ to be a function of the normal derivatives of the metric, this term is not of concern. Thus, we have succeeded in separating out a total surface derivative and a total variation (to be cancelled by adding a counter-term to the action) so that the boundary variation of the action can be put to zero by fixing the metric at the boundary, \textit{without specifying timelike or null nature of the boundary surface.} The counter-term that can be added to the Einstein-Hilbert action is the integral over the boundary of 
\begin{equation}\label{CT0}
\mathcal{C}_{0}\equiv 2 \sqrt{-g}\P[a]{b} \nabla_a s^{b}~.
\end{equation}
For null surfaces, this counter-term can be reduced to the counter-term we obtained for null surfaces in \cite{Parattu:2015gga} (given in \ref{CT01}). With $s_a=\ell_a$, the null normal, and choosing $t^a$ as an auxiliary null vector $k^a$, $\P[a]{b} \nabla_a \ell^{b}=q^{a}_{b} \nabla_a \ell^{b}-k_b \ell^a\nabla_a \ell^{b} = \Theta + \kappa$ and $\sqg=\sqrt{q}$. For non-null surfaces, the choice $t^a=-s^a / s^2$ will reduce $\mathcal{C}_{0}$ to the GHY term, as we shall prove in the next section.

\section{Connection of Our Counter-term with the Einstein Term and the GHY Term}\label{CT_compare}

Since our counter-term, the GHY counter-term and the Einstein counter-term succeed in removing the variations of the normal derivatives on the boundary, the differences between them should only involve the metric and its surface derivatives. To compare $\mathcal{C}_{0}$, $\mathcal{C}_{\textrm{GHY}}$ and $\mathcal{C}_{E}$, we shall express each of them in terms of $s_a$, $t^a$ and $g_{ab}$. From \ref{CT0}, we have
\begin{equation} \label{CT0_exp}
	\mathcal{C}_{0}= 2 \sqrt{-g}\P[a]{b} \nabla_a s^{b} = 2 \sqg \left( \df^a_b + t^a s_b \right)\nabla_a s^b=2 \sqg \nabla_a s^a + \sqg   t^a  \nabla_a s^2~.
\end{equation}
For the Gibbons-Hawking term, we write the normalized normal as $n_a=Ns_a$, with the normalization factor $N$ defined by the equation $g^{ab}s_a s_b = \epsilon/N^2$, with $\epsilon=-1/+1$ for spacelike/timelike surface. (We shall take $N$ to be positive, but a negative $N$ can be accommodated easily.) Then, $\mathcal{C}_{\textrm{GHY}}$ (\ref{CTGHY1}) may be expanded as follows:
\begin{align}\label{CTGHY_exp}
	\mathcal{C}_{\textrm{GHY}} = 2\sqh \nabla_a n^a = \frac{2\sqg}{N}\nabla_a \left( N s^a\right) = 2\sqg \nabla_a s^a + 2\sqg s^a \nabla_a \ln N~,
\end{align}
where we have used $\sqg=N \sqh$.  Finally, we expand the Einstein term (\ref{CTE1}). First, let us manipulate the expression for $V^c$.
\begin{equation}
	V^c = -\frac{1}{g} \partial_b \left(g g^{bc} \right) = -\partial_b g^{bc} - 2 g^{bc} \partial_b \ln \sqg = - \frac{\partial_b \left(\sqg g^{bc} \right)}{\sqg} - g^{bc} \frac{\partial_b \sqg}{\sqg}~.
\end{equation}
Then, substituting in the expression for $\mathcal{C}_{E}$, we obtain
\begin{align}\label{CTE_exp}
	\mathcal{C}_{E}= - \sqg s_c V^c =s_c \partial_b \left(\sqg g^{bc}\right) + s^{b} \partial_b \sqg 
	= 2 \sqg \nabla_a s^a -\sqg g^{bc}\partial_b s_c- \sqg\partial_b s^{b}~.
\end{align}
Note that the term $2 \sqg \nabla_a s^a$ is common in $\CO$, $\CG$ and $\CE$ as seen from \ref{CT0_exp}, \ref{CTGHY_exp} and \ref{CTE_exp}. But it alone won't suffice as a counter-term as the extra terms in all three counter-terms contain normal derivatives of the metric. 

Let us first compare $\CO $ and $\CG$ in the case of a non-null surface. The second term in $\CO$ is
\begin{align}
	\sqg   t^a  \partial_a s^2 &=  \sqg t^a  \partial_a \left(\frac{\epsilon}{N^2} \right)=-2 \sqg \left(\frac{\epsilon}{N^2}\right) t^a\partial_a \ln N= -2 \sqg s^2 t^a\partial_a \ln N~,   \label{normal_der}
\end{align}
where we have used $g^{ab}s_a s_b =\epsilon/N^2$. Clearly, the choice $t^a=-s^a / s^2$, which reduces $\P[a]{b}$ to $h^a_b$, will reduce this term to the second term in $\CG$ and render $\CO=\CG$. To separate the normal derivatives for a general $t^a$, we shall use our 3+1 coordinates (\ref{spec_coord}) and compare the coefficients of the $\partial_\phi \ln N$ term. Then, $ t^\phi=t^a s_a=-1$ and $s^{\phi}=s^a s_a=s^2$, proving that the normal derivatives are the same in both terms.
 
Next, let us compare $\CO$ and $\CE$. The extra terms in $\CE$ are
\begin{align}
-\sqg \left(g^{bc}\partial_b s_c + \partial_b s^{b}\right)=-\sqg \left(g^{bc}\partial_b s_c +\left(\P[a]{b}-t^a s_b \right) \partial_a s^{b}\right)
=-\sqg \left(\Pi^{ab}\partial_a s_b +\P[a]{b}\partial_a s^{b}- t^a \partial_a s^{2}\right)~.
\end{align} 
The last term is exactly the extra term in $\CO$ in \ref{CT0_exp}. The first two terms have only surface derivatives of the metric, most easily seen in our ``3+1'' coordinates (\ref{spec_coord}) as $\Pi^{\phi b}=\Pi^{a b}s_a=0$. 

Thus, we have proved that all three counter-terms in consideration, the terms with normal derivatives of the metric are common. In the 3+1 coordinates (\ref{spec_coord}), all the counter-terms we have discussed have the structure
\begin{align}
\mathcal{C}=2\sqg \nabla_a s^a + 2 \sqg s^2 \partial_\phi \ln N + \mathcal{D}, 
\end{align}
with the extra term $\mathcal{D}$ having the expressions $\mathcal{D}_{0}=- 2 \sqg s^2 t^{\alpha} \partial_\alpha \ln N$, $\mathcal{D}_{\rm GHY}=2 \sqg s^{\alpha} \partial_\alpha \ln N$ and $\mathcal{D}_{E}=- \sqg \partial_{\alpha} s^{\alpha}$ (since $\partial_a s_b=0$ in these coordinates).

\section{Results and Conclusions} 
We have introduced a boundary counter-term that can be added to the Einstein-Hilbert action without differentiating between null and non-null surfaces. Starting with a $\phi=$ constant surface as the boundary of the spacetime region in which the action is varied, we take $s_a = \partial_a \phi$ as the normal and some vector $t^a$ as the auxiliary vector satisfying $t^a s_a =-1.$ The norm of $s_a$ or $t^a$ are not specified. We could show that the boundary term on the  $\phi=$ constant surface, irrespective of whether the surface is null, spacelike or timelike, can be decomposed as
\begin{align} \label{final_result_1}
 \ST = \intdx~&\left\{  \partial_a \left(\sqg \P[a]{b}\df s^b \right) - 2 \delta (\sqrt{-g}\P[a]{b} \nabla_a s^{b}) +  \sqg \df t^a \partial_a \left(s^2\right) \right.\nn\\
 &\left.+ \sqrt{-g} \left[\nabla_a s_b -g_{ab} \left(\P[c]{d} \nabla_c s^{d}\right)- (\nabla_c t^c)s_a s_b \right] \delta g^{ab}\right\},
\end{align}
where $\P[a]{b} = \df^a_b + t^a s_b$. The first term is a total surface derivative and the second term may be cancelled by adding the integral of $2 \sqrt{-g}\P[a]{b} \nabla_a s^{b}$ over the boundary as a counter-term to the action. The rest of the terms can be eliminated by fixing the metric on the boundary. Thus, the counter-term that can be added to the Einstein-Hilbert action is the integral over the boundary of 
\begin{equation}\label{CT0-final}
\mathcal{C}_{0}\equiv 2 \sqrt{-g}\P[a]{b} \nabla_a s^{b}~.
\end{equation}
We also explicitly evaluated the difference between our counter-term $\CO$, the Gibbons-Hawking-York counter-term $\CG$ and the Einstein counter-term $\CE$, and showed that the expressions differ only by terms that do not involve normal derivatives. For non-null surfaces, we can choose $t^a=-s^a /s^2$ and reduce our counter-term to the GHY counter-term. For null surfaces, choosing $t^a=k^a$, the auxiliary null vector, reduces $\CO$ to the counter-term we proposed in a previous paper \cite{Parattu:2015gga}.

\section*{Acknowledgments}

The research of TP is partially supported by J.C.Bose research grant of DST, India. 
KP and SC are supported by the Shyama Prasad Mukherjee Fellowship from the Council of Scientific and 
Industrial Research (CSIR), India. KP and SC would like to thank Kinjalk Lochan for discussions. 


\bibliography{mybibliography-gravity}

\providecommand{\href}[2]{#2}\begingroup\raggedright\begin{thebibliography}{10}

\bibitem{Dyer:2008hb}
E.~Dyer and K.~Hinterbichler, ``{Boundary Terms, Variational Principles and
  Higher Derivative Modified Gravity},'' {\em Phys.Rev.} {\bf D79} (2009)
  024028,
\href{http://www.arXiv.org/abs/0809.4033}{{\tt 0809.4033}}.

\bibitem{York:1972sj}
J.~York, James~W., ``{Role of conformal three geometry in the dynamics of
  gravitation},'' {\em Phys.Rev.Lett.} {\bf 28} (1972)
1082--1085.

\bibitem{Gibbons:1976ue}
G.~Gibbons and S.~Hawking, ``{Action Integrals and Partition Functions in
  Quantum Gravity},'' {\em Phys.Rev.} {\bf D15} (1977)
2752--2756.

\bibitem{Charap:1982kn}
J.~Charap and J.~Nelson, ``{Surface Integrals and the Gravitational Action},''
  {\em J.Phys.A:Math.Gen.} {\bf 16} (1983)
1661.

\bibitem{Einstein:1916cd}
A.~Einstein, ``{Hamilton's Principle and the General Theory of Relativity},''
  {\em Sitzungsber.Preuss.Akad.Wiss.Berlin (Math.Phys.)} {\bf 1916} (1916)
1111--1116.

\bibitem{Parattu:2015gga}
K.~Parattu, S.~Chakraborty, B.~R. Majhi, and T.~Padmanabhan, ``{Null Surfaces:
  Counter-term for the Action Principle and the Characterization of the
  Gravitational Degrees of Freedom},''
\href{http://www.arXiv.org/abs/1501.01053}{{\tt 1501.01053}}.

\bibitem{gravitation}
T.Padmanabhan, {\em {Gravitation: Foundations and Frontiers}}.
\newblock Cambridge University Press, Cambridge, UK, 2010.

\bibitem{Arnowitt:1962hi}
R.~L. Arnowitt, S.~Deser, and C.~W. Misner, ``{The Dynamics of general
  relativity},'' \href{http://www.arXiv.org/abs/gr-qc/0405109}{{\tt
  gr-qc/0405109}}.

\bibitem{Parattu:2013gwb}
K.~Parattu, B.~R. Majhi, and T.~Padmanabhan, ``{Structure of the gravitational
  action and its relation with horizon thermodynamics and emergent gravity
  paradigm},'' {\em Phys.Rev.} {\bf D87} (2013), no.~12, 124011,
\href{http://www.arXiv.org/abs/1303.1535}{{\tt 1303.1535}}.

\bibitem{Wald}
R.~M. Wald, {\em {General Relativity}}.
\newblock The University of Chicago Press, 1st~ed., 1984.

\bibitem{Poisson}
E.~Poisson, {\em {A Relativist's Toolkit: The Mathematics of Black-Hole
  Mechanics}}.
\newblock Cambridge University Press, 1st~ed., 2007.

\bibitem{Hawking:2010mca}
S.~W. Hawking and W.~Israel~eds, {\em {General Relativity: An Einstein
  Centenary Survey}}.
\newblock Cambridge University Press, 1979.

\bibitem{Padmanabhan:2014BT}
T.~Padmanabhan, ``{A short note on the boundary term for the Hilbert action},''
  {\em Mod.Phys.Lett.} {\bf A29} (2014)
1450037.

\end{thebibliography}\endgroup

\bibliographystyle{./utphys}

\end{document}